\documentclass[a4paper,12pt]{amsart}

\usepackage[]{amsmath}
\usepackage{amsfonts}
\usepackage{amssymb}
\usepackage{xcolor}
\usepackage{mathrsfs}
\usepackage[numbers,sort]{natbib}
\usepackage{natbib}

\usepackage{tikz}
\usetikzlibrary{shapes.geometric,positioning,calc,cd}
\usepackage{binarytree} 
\usepackage[margin=3cm]{geometry} 
\usepackage{hyperref}

\usepackage{algorithm}
\usepackage{algpseudocode}

\newtheorem{thm}{Theorem}[section]
\newtheorem{prop}[thm]{Proposition}
\newtheorem{lem}[thm]{Lemma}

\theoremstyle{definition}

\newtheorem{rem}[thm]{Remark}

\usepackage{amsfonts}
\usepackage{graphicx} 
\newcommand{\C}{\mathcal C}

\newcommand{\N}{\mathcal N}
\newcommand{\BN}{\mathbb N}
\newcommand{\T}{\mathcal T}
\newcommand{\PP}{\mathbb P}

\title[`Tight' classes of phylogenetic networks]{A dichotomy law for certain classes of phylogenetic networks}

\author{Michael Fuchs and Mike Steel}
\date{\today}

\begin{document}
\begin{abstract}
Many classes of phylogenetic networks have been proposed in the literature. A  feature of several of these classes is that if one restricts a network in the class  to a subset of its leaves, then the resulting network may no longer lie within this class. This has implications for their biological applicability, since some species -- which are the leaves of an underlying evolutionary network -- may be missing (e.g., they may have become extinct, or there are no data available for them) or we may simply wish to focus attention on a subset of the species.  On the other hand, certain classes of networks are `closed' when we restrict to subsets of leaves, such as  (i) the classes of all phylogenetic networks or all phylogenetic trees; (ii) the classes of galled networks, simplicial networks, galled trees; and (iii) the classes of networks that have some parameter that is monotone-under-leaf-subsampling (e.g., the number of reticulations, height, etc.) bounded by some fixed value. It is easily shown that a closed subclass of phylogenetic trees is either all trees or a vanishingly small proportion of them (as the number of leaves grows). In this short paper, we explore whether this dichotomy phenomenon holds for other classes of phylogenetic networks, and their subclasses.
\end{abstract}
\maketitle

\section{Introduction}

Phylogenetic networks provide a way for biologists to represent the evolutionary history of present-day species more accurately than traditional phylogenetic trees allow. Whereas trees can adequately represent past speciation and extinction events, networks can also explicitly display reticulate evolutionary events, such as lateral gene transfer and hybridization \cite{bap13, hus11}.  The leaves (`tips') of these networks are typically a set of present-day species, and the root of the network represents their most recent common ancestor.  Over the last two decades, the definition and study of various phylogenetic classes, their enumeration and combinatorial properties, and deciphering the relationships between these various classes has been an active area of research \cite{kon}.

A desirable property of any class of networks is that it satisties a certain `closure' property. Roughly speaking, this property states that if a network in the class is restricted to a subset of its leaf set, then the induced sub-network remains in that class. 
For example, the class of phylogenetic trees has this property, as does the class of all phylogenetic networks. However, many other classes between these two extremes  (e.g., tree-child networks, tree-based networks, etc.) fail to have this closure property.

If a network class is not closed, it might be hoped that a large closed subclass exists within it, perhaps even one that might even be asymptotically equivalent in size to the full class, as the number of leaves grow.   However, in this paper, we demonstrate that for certain classes of networks  this is far from the case: for certain `tight' classes of networks, every closed subset of the class is either the entire class, or it constitutes a vanishingly small proportion of the entire class (as the number of leaves  becomes large). Our methods rely primarily on asymptotic enumeration techniques.  We end by posing a general question for further study.

\subsection{Definitions: Phylogenetic networks}\label{5-classes}
\begin{figure}[htb]
\centering
\includegraphics[scale=0.63]{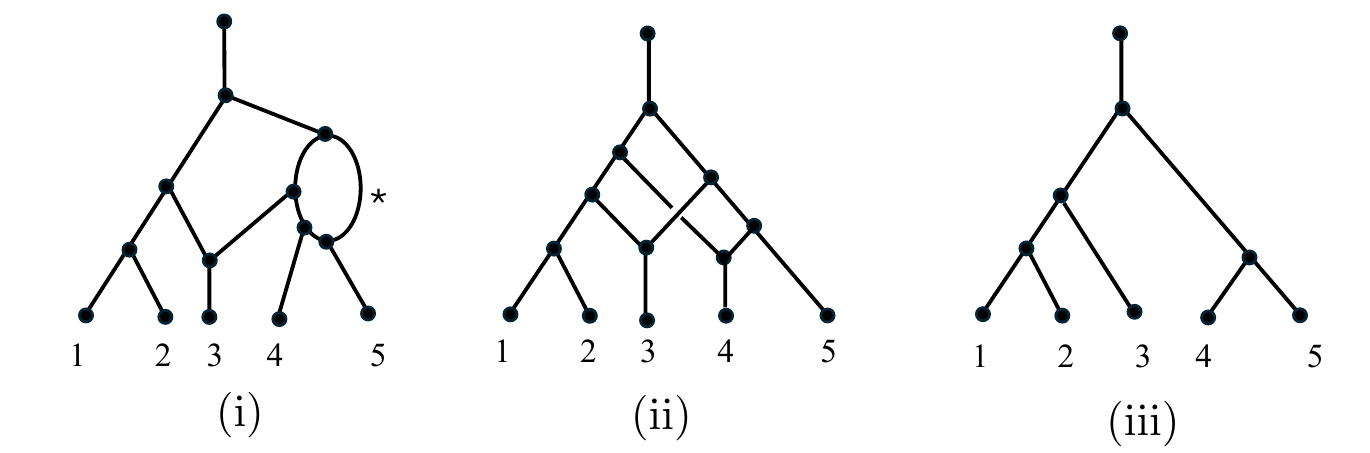}
\caption{Three phylogenetic networks, each having leaf set $\{1,2,3,4,5\}$. Edges are directed downwards. (i) A tree-child network,  (ii) a normal network, and (iii) a phylogenetic tree. Note that although the networks in (i) and (ii) each have two reticulation vertices, the network in (i) is not a normal network due to the presence of a `shortcut' edge indicated by *.}
\label{fig1}
\end{figure}

Throughout this paper,  all trees and networks are directed graphs, represented by a set of vertices and a set of edges (ordered pairs of distinct vertices). We now recall some standard terminology in the phylogenetic literature. 
  A (binary) {\em phylogenetic network} on $[n]:=\{1,\ldots,n\}$ is a directed acyclic graph with $n$ leaves (vertices of out-degree 0) labeled bijectively by the elements of $[n]$, and with each non-leaf vertex having in-degree 1 and out-degree 2 (tree vertices), or in-degree 2 and out-degree 1 (reticulation vertices or reticulations for short), or in-degree 0 and out-degree 1 (the root of the network at the top of an ancestral root edge). 
Edges which terminate in a reticulation vertex are called {\it reticulation edges}; all other edges are called {\it tree edges}. Two phylogenetic networks are regarded as equivalent if there is a directed graph isomorphism between them that maps $i$ to $i$ for each $i \in [n]$.  Three important classes of phylogenetic networks are the following:
\begin{itemize}
    \item A {\em tree-child network}  is a phylogenetic network for which each non-leaf vertex has at least one of its outgoing edges directed to a tree vertex or a leaf.
    \item A {\em normal network} is a tree-child network that has no `shortcut' edge (i.e., no edge $(u,v)$ for which there is another path from $u$ to $v$).
        \item A {\em phylogenetic tree} is a phylogenetic network with no reticulation vertices.
\end{itemize}
Thus, tree-child networks include normal networks which, in turn, include phylogenetic trees.
For more background and details on phylogenetic networks; see \cite{hus11}.
Fig.~\ref{fig1} illustrates these three classes of networks.

\begin{figure}[htb]
\centering
\includegraphics[scale=0.7]{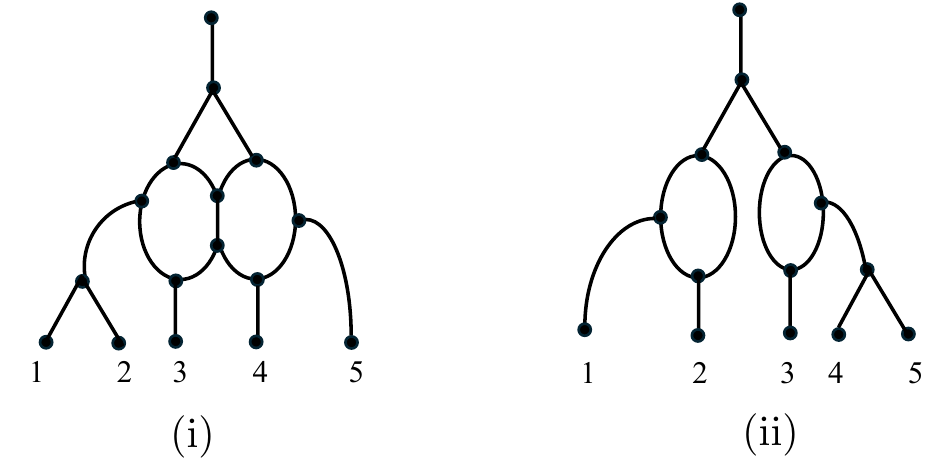}
\caption{(i) A galled network, (ii) a galled tree. Edges are directed downwards.}
\label{fig2}
\end{figure}

Apart from the above three classes of phylogenetic networks, two more will play an important role in this paper. For the definition of these two, we need the notion of a {\it reticulation cycle} (or {\it gall}) which is a set of two paths from a common top tree vertex to a common bottom reticulation, with the sets of internal vertices being disjoint. Then, we have the following definitions:
\begin{itemize}
\item A {\it galled network} is a phylogenetic network with each reticulation in exactly one reticulation cycle.
\item A {\it galled tree} is a galled network whose reticulation cycles are edge-disjoint.
\end{itemize}
Fig.~\ref{fig2} illustrates these two classes of networks.

For more types of networks, we refer the reader to \cite{hus11}.

\subsection{Definitions: Classes of networks} A {\em class of networks} $\N$ is the set of all binary networks of a particular type  (e.g., tree-child, normal, trees, etc.), and having a leaf set that is a subset of $\BN=\{1,2,3,\ldots\}$. We will also assume that any class of networks $\N$ satisfies the following two properties:

\begin{itemize}
    \item[($P_1$)] If $N \in \N$, then any relabelling of the leaves of $N$ by distinct elements of $\BN$ results in a network that also lies in $\N$\footnote{Note that because of this property, a class of networks is not necessary a combinatorial class (see, e.g., Definition I.6 in \cite{FlSe}) as the number of networks with a fixed number of leaves is infinite. However, subsequently, we will restrict to sets of networks whose leaf sets are a subset of $[n]$ and these sets will have a finite cardinality.}.
    \item[ ($P_2$)] For some fixed function $f$, the number of vertices in any network in $\N$ is bounded above by $f(n)$, where $n$ is the number of leaves of $N$.
\end{itemize}

Property ($P_1$) captures the requirement that classes of networks depend only on the shape of the network and not on the way its leaves are labelled. Property $(P_2)$  is satisfied by many of the well-studied classes of phylogenetic networks (e.g., trees, tree-child networks, galled networks, etc.). In addition, sufficient conditions are known so that a class satisfies ($P_2$); see \cite{Se}. However,  $(P_2)$ excludes, for example, the class of {\em all} phylogenetic networks, or all tree-based networks, since such networks can have an unbounded number of vertices and yet have just two leaves.  

For a class of networks $\N$ and a subset $X$ of $\BN$, let $\N(X)$ be the set of networks in $\N$ that have leaf set $X$ and let
$\N_n = \bigcup_{X \subseteq [n]}\N(X)$.
Notice that $\N_n \subseteq \N_{n+1}$ and $$\mathcal{N}= \bigcup_{n \geq 1} \mathcal{N}_n,$$ 
(i.e., this union equals $\N$ as leaf sets of the networks in $\N$ may be arbitrary subsets of ${\mathbb N}=\{1,2,3,\ldots\}$).

\begin{figure}[htb]
\centering
\includegraphics[scale=0.53]{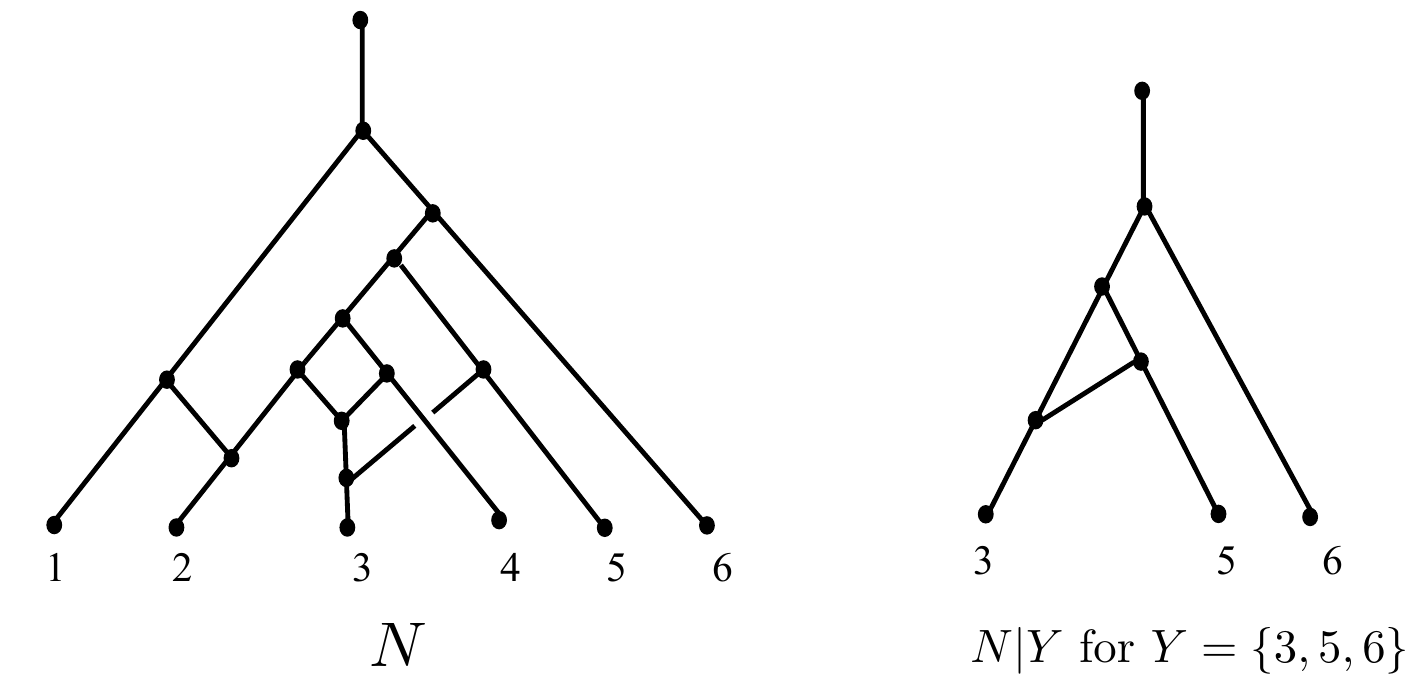}
\caption{Restricting a network to a subset of leaves. Edges are directed downwards.}
\label{fig3}
\end{figure}

For  $N \in \N_n$, let ${\mathcal L}_N$ denote the set of leaves of $N$ (a subset of $[n]=\{1, 2, \ldots, n\}$) and, given a subset $Y$ of ${\mathcal L}_N$, let $N|Y$ be the induced phylogenetic network on the restricted leaf set $Y$. The network $N|Y$ is obtained from $N$ by taking all vertices and edges of $N$ that lie on at least one path from the root of $N$ to at least one leaf in $Y$, and then suppressing any resulting subdivision vertices (i.e. vertices of in-degree and out-degree equal to 1) and any double edges. Fig.~\ref{fig3} illustrates this notion, and involves both types of suppression.

\subsection{Closed subsets of $\N$ and closed classes}

A subset ${\mathcal C}$ of $\N$ is said to be {\em closed} if $\C$ satisfies properties $(P_1), (P_2)$ and a further property:
\begin{itemize}
    \item[($P_3$)] If $N \in \C$, and $Y \subseteq {\mathcal L}_N$ then $N|Y \in \C.$
\end{itemize}
If $\C=\N$, then the class itself is called closed. Examples of closed classes of networks include the following.
\begin{itemize}
    \item The class of all phylogenetic trees (or any closed subclass, as discussed in the next section);
    \item Galled trees (i.e., level-1 networks);
    \item Galled networks;
   \item 
   The class of networks with at most $k$ reticulation vertices (for any $k\geq 1$);
    \item The class of networks of height at most $k$ (for any $k\geq 1$);
    \item An arbitrary union of two or more closed classes;
    \item The class of {\em simplicial} networks --- i.e., networks for which the child of every reticulation is a leaf;
    \item The class of {\em semi-simplicial} networks --- i.e., networks for which the child of every reticulation is either a leaf or the root of a tree;
    \item The class of networks defined by $f(N)\leq k$, where $f$ is a function that satisfies $f(N|Y) \leq f(N)$ for all $Y \subseteq {\mathcal L}_N$  and $k$ is a fixed value (this example generalises some of the entries above). 
\end{itemize}
Examples of classes of networks that are not closed (because they fail to satisfy Properties ($P_2$) or ($P_3$) or both) include tree-child networks, normal networks, tree-based networks, and orchard networks. (See \cite{kon} for a definition of the latter two network classes.)

\section{Closed classes of Phylogenetic trees}\label{pts}

Here, `tree' refers to any finite rooted binary phylogenetic tree.
Given any subset $X$ of $[n]$, let $\T(X)$ be the set of all trees on leaf set $X$. Thus $|\T(X)|= (2|X|-3)!!:=(2|X|-3)(2|X|-5)\cdots 1$ (where the empty product is by convention equal to one); see, e.g., Corollary 2.2.4 in \cite{SeSt}. Next, let $$\T_n = \bigcup_{X \subseteq [n]} \T(X).$$ 
Notice that $\T_n \subseteq \T_{n+1}$  and
\[
\vert\T_n\vert=\sum_{j=1}^{n}\binom{n}{j}(2j-3)!!=n!\sum_{j=1}^{n}\frac{C_{j-1}}{(n-j)!2^{j-1}}=\frac{n!}{2^{n-1}}\sum_{j=0}^{n-1}\frac{2^{j}C_{n-j-1}}{j!},
\]
where $C_n=\frac{1}{n+1}\binom{2n}{n}$ denotes the Catalan numbers, which have the asymptotic expansion
\[
C_{n-j-1}=\frac{4^{n-j-1}}{\sqrt{\pi n^3}}\left(1+{\mathcal O}\left(\frac{j+1}{n}\right)\right)
\]
uniformly for $j=o(n)$,  as $n\rightarrow\infty$. In addition, by considering the quotient of consecutive terms of the sequence $2^jC_{n-j-1}/j!, 0\leq j\leq n-1$, it can be seen that the sequence is decreasing. Thus, by an application of the Laplace method (for a detailed explanation of this method, see Section 4.7 in \cite{FlSe}):
\begin{equation}\label{asym-Tn}
\vert\T_n\vert\sim\sqrt{\frac{e}{\pi n^3}}2^{n-1}n!\sim\sqrt{\frac{e}{2}}n^{-1}\left(\frac{2}{e}\right)^nn^{n},\qquad (n\rightarrow\infty),
\end{equation}
where we used Stirling's formula in the last step.

We now consider the infinite set $$\T = \bigcup_{n \geq 1} \T_n,$$ 
which is the set of {\em all} finite phylogenetic trees whose leaf set consists of an arbitrary finite subset of $\BN = \{1,2,3, \ldots \}.$

Examples of closed subsets of 
$\T$ include:

\begin{itemize}
\item $\T$;
\item The class of caterpillar trees, i.e., trees with exactly one cherry;
\item The class of trees that have height at most $k$  (for any $k \geq 1$);
\item The class of trees that have at most $k$ cherries  (i.e., pairs of leaves incident with a common vertex) for any $k \geq 1$;
\item For a fixed tree, the class of leaf relabellings, and all of the induced subtrees of these trees; 
\item An arbitrary union of two or more closed subsets of $\T.$
\end{itemize}

Our first result states that a closed class of trees is either all trees, or an asymptotically negligible proportion of trees.

\begin{prop}
\label{pro1}
      Let $\C$ be a closed subset of $\T$.  Then either $\C =\T$ or 
$$\lim_{n \rightarrow \infty} \frac{|\C \cap \T_n|}{|\T_n|} = 0.$$  
\end{prop}

\begin{proof}
   If $\C \neq \T$, then there exists a tree $T\in \T$ that is not present in $\C$. Let $\tau$ be the tree shape of $T$, i.e., $T$ with the labels of the leaves removed. Since $\C$ is closed, none of the trees in $\C$ contains a subtree of shape $\tau$. 
   Let $T_n$ be a tree sampled uniformly at random from $\T_n$. We show that
   $\PP(T_n \in \C) \rightarrow 0$ as $n \rightarrow \infty.$
Let  $E_n$ be the event that $T_n$ contains a subtree of shape $\tau$ (this subtree shape can be anywhere inside $T_n$).
By the law of total probability, we have
\begin{equation}
\label{eqx}
      \PP(E_n) =\PP(E_n|T_n \in \C)\PP(T_n \in \C) +\PP(E_n|T_n \not\in \C)\PP(T_n \not\in \C).
\end{equation}
Now, $\PP(E_n|T_n \in \C)=0$ (by definition), and the second (product) term on the right of Eqn.~\ref{eqx} is less than or equal to $\PP(T_n \not\in \C)$. Thus $\PP(E_n) \leq \PP(T_n \not\in \C)$.
We now apply Corollary 13 of \cite{bie}, which imples that $\PP(E_n)$ tends to 1 as $n \rightarrow \infty$. Consequently, as $n \rightarrow \infty$, we have $\PP(T_n \not\in \C)   \rightarrow 1$ and thus $\PP(T_n \in \C) =\frac{|\C \cap \T_n|}{|\T_n|} \rightarrow 0$.
\end{proof}

\section{Extending Proposition~\ref{pro1} beyond trees}\label{gen-pn}

For the remainder of this paper we explore the extent to which Proposition~\ref{pro1}  holds for more general classes of phylogenetic networks. We begin by formalizing this notion.
\subsection{Definitions: Tightness} We say that a class of networks $\N$ that satisfies ($P_1, P_2)$ is {\em tight} if the following dichotomy condition holds.
    For every closed subset  $\C$ of $\N$, either $\C =\N$ or 
  \begin{equation}
      \label{zero}
      \lim_{n \rightarrow \infty} \frac{|\C \cap \N_n|}{|\N_n|} = 0. 
  \end{equation}

\bigskip


Notice that if $\N$ is {\bf not} a closed class of networks (i.e., it violates condition ($P_3$)), then the condition that $\N$ is tight is equivalent to the statement that, for any closed subset $\C$ of $\N$, Eqn.~\ref{zero} holds.  

Notice also that every class of networks $\N$ contains a unique maximal closed class with respect to set inclusion, namely, the union of {\em all} closed classes in $\N$ (which, as noted above, forms a closed class). We denote this maximal closed class for $\N$ by $\C_{\N}$.  Thus, if $\N\ne\C_{\N}$ (i.e., $\N$ is not closed), then $\N$ is tight if and only if Eqn.~\ref{zero} holds for the single closed class $\C=\C_{\N}$.

\subsection{Examples} We give some examples of classes which are tight and not tight

{\bf Example 1:} The set $\T$ of all binary phylogenetic trees is tight, by Proposition~\ref{pro1}.

{\bf Example 2:}
Let $\N$ be the set of {\bf all} galled trees.  Then, $\N$ is tight by Corollary 14 of \cite{bie} and a similar argument to that used in Proposition~\ref{pro1}. 

{\bf Example 3:} Let $\N$ be the class of galled networks. Then $\N$ is not tight, since if we take $\C$ to be the closed subclass of simplicial galled networks, we have $\C \subsetneq \N$ but  
Eqn.~\ref{zero} does not hold due to results in \cite{fuc22}.  More precisely, we have the following.

 \begin{prop}\label{galled} If $\N$ is the class of galled networks and $\C$ is the class of simplicial galled networks, then
 \[
 \lim_{n\rightarrow\infty}\frac{\vert\C\cap\N_n\vert}{\vert\N_n\vert}=e^{-3/8}.
 \]
 \end{prop}
 \begin{proof}
 First, note that the class of simplicial galled networks and the class of all simplicial networks coincide (see, for example,  Proposition~2 in \cite{CaZh}). Next, it was proved in \cite{fuc22} that the numbers of galled networks (${\rm GN}_\ell$) and the number of simplicial networks (${\rm SN}_{\ell}$) with $\ell$ leaves that are labelled by the set $X=\{1,\ldots,\ell\}$ admit the following asymptotics, as $\ell\rightarrow\infty$,
 \begin{equation}\label{asymp-GNl}
 {\rm GN}_{\ell}\sim\frac{\sqrt{2e\sqrt[4]{e}}}{4}\ell^{-1}\left(\frac{8}{e^2}\right)^{\ell}\ell^{2\ell}\sim\frac{\sqrt{2e\sqrt[4]{e}}}{8\pi}\ell^{-2}8^{\ell}\ell!^2
 \end{equation}
 and
 \begin{equation}\label{asymp-SNl}
 {\rm SN}_{\ell}\sim\frac{\sqrt{2\sqrt{e}}}{4}\ell^{-1}\left(\frac{8}{e^2}\right)^{\ell}\ell^{2\ell}\sim\frac{\sqrt{2e\sqrt{e}}}{8\pi}\ell^{-2}8^{\ell}\ell!^2.
 \end{equation}
 We now use a similar line of reasoning as for Eqn.~\ref{asym-Tn} in Section~\ref{pts} to find the asymptotics of $\vert{\C}\cap{\N}_n\vert$ and $\vert{\N}_n\vert$. First, 
 \[
 \vert\N_n\vert=\sum_{j=1}^{n}\binom{n}{j}{\rm GN}_j={\rm GN}_n+\sum_{j=1}^{n-1}\binom{n}{j}{\rm GN}_j
 \]
 and from Eqn.~\ref{asymp-GNl}, we obtain for the sum-term:
 \[
 \sum_{j=1}^{n-1}\binom{n}{j}{\rm GN}_j={\mathcal O}\left(\sum_{j=1}^{n-1}\binom{n}{j}j^{-2}8^jj!^2\right)={\mathcal O}\left(8^nn!\sum_{j=1}^{n-1}\frac{(n-j-1)!}{j!8^j(n-j)}\right),
 \]
 where the summation index was changed from $j$ to $n-j$ in the last step. Next, observe that, as $n\rightarrow\infty,$ 
 \[
 \frac{(n-j-1)!}{j!8^j(n-j)}= \frac{1}{j!8^j}n^{-j-2}n!\left(1+{\mathcal O}\left(\frac{j^2}{n}\right)\right)
 \]
 uniformly for $j=o(\sqrt{n})$. Moreover, $(n-j-1)!/(j!8^j(n-j)),1\leq j\leq n-1$ is a decreasing sequence. Thus, by applying the Laplace method:
 \[
 \sum_{j=1}^{n-1}\binom{n}{j}{\rm GN}_j={\mathcal O}\left(n^{-3}8^nn!^2\right).
 \]
 Consequently,
 \[
 \vert\N_n\vert\sim{\rm GN}_n\sim\frac{\sqrt{2e\sqrt[4]{e}}}{4}n^{-1}\left(\frac{8}{e^2}\right)^{n}n^{2n},\qquad (n\rightarrow\infty).
 \]
 Likewise, from Eqn.~\ref{asymp-SNl},
 \[
 \vert\C\cap\N_n\vert\sim\frac{\sqrt{2\sqrt{e}}}{4}n^{-1}\left(\frac{8}{e^2}\right)^{n}n^{2n},\qquad (n\rightarrow\infty).
 \]
 The claimed result follows from these two asymptotic relations.
 \end{proof}

{\bf Example 4:} Let $\N$ be the class of semi-simplicial networks. Then $\N$ is not tight, since the closed subclass $\C$ of simplicial networks violates Eqn.~\ref{zero}, as this limit cannot tend to zero due to the previous example and the fact that semi-simplicial networks and  simplicial networks are both galled networks. In fact, we can even compute the limit precisely.
\begin{prop}
If $\N$ is the class of semi-simplicial networks and $\C$ the class of simplicial networks, then
 \[
 \lim_{n\rightarrow\infty}\frac{\vert\C\cap\N_n\vert}{\vert\N_n\vert}=e^{-1/16}.
 \]    
\end{prop}
This is proved in a similar way to Proposition~\ref{galled} by using the following result, which will be established in the appendix.
\begin{thm}
\label{ssn} The number ${\rm SSN}_{\ell}$ of semi-simplicial networks with  $\ell$ leaves that are labelled by the set $X=\{1,\ldots,\ell\}$ admits the asymptotics:
\[
{\rm SSN}_{\ell}\sim\frac{\sqrt{2\sqrt{e\sqrt[4]{e}}}}{4}\ell^{-1}\left(\frac{8}{e^2}\right)^{\ell}\ell^{2\ell},\qquad (\ell\rightarrow\infty).
\]
\end{thm}

\section{Tree-child and normal networks}

In this section, we show that (i) the classes of all tree-child and all normal networks are both tight, and (ii) the class of tree-child networks with at most $k\geq 1$ reticulations is not tight, whereas the class of normal networks with at most $k\geq 1$ reticulations is tight. 

\subsection{The classes of all tree-child and all normal networks}

First, let $\N$ be the class of all binary tree-child networks. Observe that the type of argument used to establish the tightness of the class of binary phylogenetic trees (Proposition~\ref{pro1}) cannot be used to show that $\N$ is tight, since there are tree-child networks (and indeed trees) that have a small probability of being present in a large uniformly sampled tree-child network.  To see why, observe that by
Corollary 1.10(i) of \cite{cha24}, the expected number of cherries (i.e., pairs of leaves incident with a common vertex) is ${\mathcal O}(1)$. If we now take $T$ to be a complete balanced binary tree with $2^k$ leaves (and so $2^{k-1}$ cherries), then the probability that a uniformly sampled network $N$ in $\N$ with $n$ leaves contains $T$ as a subtree (somewhere within $N$) does not tend to 1 as $n$ grows, provided that $k$ is chosen sufficiently large.

Nevertheless, we now provide a different argument to establish the following result.

\begin{thm}\label{TC-tight}
    The class of all tree-child networks is tight. 
\end{thm}

For the proof of this result, we need the following lemma (which uses the definition of a {\it reticulation cycle} from Section~\ref{5-classes}).

\begin{lem}\label{largest-TC}
The largest closed subclass $\C_{\N}$ of the class $\N$ of tree-child networks is the class of galled trees.
\end{lem}
\begin{proof} Let $\C$ be a subclass of the class of tree-child networks which contains at least one network $N$ that is not a galled tree. Then, $N$ contains a reticulation $v$ which is the bottom vertex of a reticulation cycle that contains either (i) another reticulation or (ii) a tree vertex $u$ whose child which is not on the cycle also has a reticulation as descendant. Choose the lowest such vertex on one of the paths from the bottom tree vertex of the reticulation cycle which contains $v$. Also, in Case (i), pick a leaf $\tilde{u}$ which can be reached from $v$ via a path consisting only of tree vertices (such a path exists due to the tree-child property). Then, the induced subnetwork $N|\{\tilde{u}\}$ is not tree-child (since it contains a reticulation whose child is another reticulation). In Case (ii), let $\tilde{v}$ be a reticulation on the path from $\tilde{u}$ which does not contain $v$ so that all other vertices of this path are tree nodes. Moreover, choose a leaf $\hat{u}$ which again can be reached via a path consisting only of tree vertices from $\tilde{v}$. Then, $N\vert\{\tilde{u},\hat{u}\}$ is not tree-child (as it contains a tree vertex with both children reticulations). Thus, in both cases $\C$ is not closed. This proves the desired result.
\end{proof}

\begin{proof}[Proof of Theorem~\ref{TC-tight}]
First, recall from \cite{mcd} that the number of tree-child networks (${\rm TC}_{\ell}$) with $\ell$ leaves that are labelled by the set $X=\{1,\ldots,\ell\}$ is bounded from below by
\begin{equation}\label{lb-TC}
{\rm TC}_{\ell}=\Omega(c^{\ell}\ell^{2\ell})
\end{equation}
for some constant $c>0$. Consequently,
\begin{equation}\label{lb-Nn}
\vert\N_n\vert=\sum_{j=1}^{n}\binom{n}{j}{\rm TC}_j=\Omega\left(c^n n^{2n}\right).
\end{equation}
Next, let $\C$ be the class of galled trees. Recall from \cite{BoGaMa} that the number of galled trees $({\rm GT}_{\ell})$ with $\ell$ leaves that are labelled by the set $X=\{1,\ldots,\ell\}$ is given by:
\begin{align*}
{\rm GT}_{\ell}&=\frac{\sqrt{34}(\sqrt{17}-1)}{136}\ell^{-1}\left(\frac{8}{e}\right)^{\ell}\ell^{\ell}(1+{\mathcal O}(\ell^{-1}))\\
&=\frac{\sqrt{17}(\sqrt{17}-1)}{136\sqrt{\pi}}\ell^{-3/2}8^{\ell}\ell!(1+{\mathcal O}(\ell^{-1})),\qquad (\ell\rightarrow\infty).
\end{align*}
Thus, by the Laplace method:
\begin{align*}
\vert\C\cap\N_n\vert=\sum_{j=1}^{n}\binom{n}{j}{\rm GT}_j&=\frac{\sqrt{17}(\sqrt{17}-1)n!}{136\sqrt{\pi}}\sum_{j=1}^{n}\frac{j^{-3/2}8^j}{(n-j)!}(1+{\mathcal O}(j^{-1}))\\
&\sim\frac{\sqrt{17\sqrt[4]e}(\sqrt{17}-1)}{136\sqrt{\pi}}n^{-3/2}8^nn!\\
&\sim\frac{\sqrt{34\sqrt[4]e}(\sqrt{17}-1)}{136}n^{-1}\left(\frac{8}{e}\right)^{n}n^{n},\qquad (n\rightarrow\infty).
\end{align*}
Consequently, $\C$ satisfies Eqn.~\ref{zero} and from the statement in the third paragraph of Section~\ref{gen-pn}, we obtain that $\N$ is tight as claimed.
\end{proof}
\begin{rem}
In \cite{fuc}, Eqn.~\ref{lb-TC} was refined to
\[
{\rm TC}_{\ell}=\Theta\left(\ell^{-2/3}e^{a_1(3\ell)^{1/3}}\left(\frac{12}{e^2}\right)^{\ell}\ell^{2\ell}\right),\qquad (\ell\rightarrow\infty),
\]
where $a_1$ denotes the largest root of the Airy function of the first kind. Using this result, Eqn.~\ref{lb-Nn} can also be refined as follows
\[
\vert \N_n\vert=\Theta\left(n^{-2/3}e^{a_1(3 n)^{1/3}}\left(\frac{12}{e^2}\right)^{n}n^{2n}\right),\qquad (n\rightarrow\infty).
\]
\end{rem}

Now, let us turn to the class $\N$ of normal networks. Here, similar to Lemma~\ref{largest-TC}, we have the following result.

\begin{lem}\label{largest-N}
The largest closed subclass $\C_{\N}$ of the class $\N$ of normal networks is the class of phylogenetic trees.
\end{lem}
\begin{proof}
Since every normal network is also tree-child, Lemma~\ref{largest-TC} implies that $\C_{\N}$ must be contained in the class of galled trees. Assume now that $\C_{\N}$ contains a network $N$ which is not a phylogenetic tree. Then, $N$ contains a reticulation $v$ which is in a reticulation cycle whose vertices apart from $v$ are all tree vertices. Moreover, since all networks are simple, there is at least one tree vertex $\tilde{v}$ which is not the top tree vertex of this reticulation cycle. Let $u$ be a leaf which can be reached via a path that only contains tree vertices from $v$ and let $\tilde{u}$ be a leaf which also can be reached via a path that only contains tree vertices from $\tilde{v}$. Then, the network $N\vert\{u,\tilde{u}\}$ is not normal. This is a contradiction, since $\C_{\N}$ was assumed to be closed. Thus, $\C_{\N}$ must be the class of phylogenetic trees.
\end{proof}

We now apply this lemma to show that $\N$ is tight as well.

\begin{thm}
The class of all normal networks is tight.
\end{thm}

\begin{proof}
From \cite{mcd}, we know that the number of normal networks $({\rm NN}_{\ell})$ with $\ell$ leaves that are labelled by the set $X=\{1,\ldots,\ell\}$ is bounded from below by
\[
{\rm NN}_{\ell}=\Omega(c^{\ell}\ell^{2\ell})
\]
for some constant $c>0$. Thus,
\[
\vert\N_n\vert=\sum_{j=1}^{n}\binom{n}{j}{\rm NN}_j=\Omega(c^n n^{2n}).
\]
Moreover, for the largest closed class $\C_{\N}$ in $\N$, from Lemma~\ref{largest-N} and Eqn.~\ref{asym-Tn} we have,
\[
\vert\C_{\N}\cap\N_n\vert=\vert\T_n\vert\sim\sqrt{\frac{e}{2}}n^{-1}\left(\frac{2}{e}\right)^nn^{n},\qquad (n\rightarrow\infty).
\]
Consequently, Eqn.~\ref{zero} holds and thus the claim is proved.
\end{proof}

\subsection{Tree-child and normal networks with a bounded number of reticulations}

Now, suppose that $\N$ is the class of tree-child networks with at most $k$ (fixed) reticulations where $k\geq 1$.  We are going to establish the following.

\begin{prop}
The class $\N$ of tree-child networks with at most $k$ reticulations is not tight for each $k\geq 1$.
\end{prop}
\begin{proof}
We deal with the cases $k\geq 2$ and $k=1$ separately.

For the case $k \geq 2$, it  was shown in \cite{FuHuYu} that the number of tree-child networks $({\rm TC}_{\ell,k})$ with exactly $k$ reticulations and $\ell$ leaves that are labelled by the set $X=\{1,\ldots,\ell\}$ satisfies, as $\ell\rightarrow\infty$,
 \[
{\rm TC}_{\ell,k}=\frac{2^{k-1}\sqrt{2}}{k!}\ell^{2k-1}\left(\frac{2}{e}\right)^{\ell}\ell^{\ell}\left(1+{\mathcal O}(\ell^{-1/2})\right)=\frac{2^{k-1}}{k!\sqrt{\pi}}2^{\ell}\ell^{2k-3/2}\ell!\left(1+{\mathcal O}(\ell^{-1/2})\right).
 \]
Thus, the number of tree-child networks with at most $k$ reticulations (denoted by ${\rm TC}_{\ell,\leq k}$) also satisfies the same asymptotic result, since:
\[
{\rm TC}_{\ell,\leq k}=\sum_{i=0}^{k}{\rm TC}_{\ell,i}.
\]
Combining the last two equations gives:
\[
\vert\N_{n}\vert=\sum_{j=1}^{n}\binom{n}{j}{\rm TC}_{j,\leq k}=\frac{2^{k-1}n!}{k!\sqrt{\pi}}\sum_{j=1}^{n}\frac{2^jj^{2k-3/2}}{(n-j)!}\left(1+{\mathcal O}(j^{-1/2})\right),
\]
and using the Laplace method, we obtain
\[
\vert\N_n\vert\sim\frac{2^{k-1}}{k!}\sqrt{\frac{e}{\pi}}n^{2k-3/2}2^nn!\sim\frac{2^{k-1}\sqrt{2e}}{k!}n^{2k-1}\left(\frac{2}{e}\right)^n n^{n}.
\]

Next, by the proof of Lemma~\ref{largest-TC}, the largest closed subclass of $\N$ is the class of galled trees with at most $k$ reticulations. Denote this subclass by $\C$. It was shown in \cite{AgFuGiRo} that the number of galled trees (${\rm GT}_{\ell,k}$) with exactly $k$ reticulations and $\ell$ leaves that are labelled by the set $X=\{1,\ldots,\ell\}$ satisfies, as $\ell\rightarrow\infty$, 
\[
{\rm GT}_{\ell,k}=\frac{2^{2k-1}\sqrt{2}}{(2k)!}\ell^{2k-1}\left(\frac{2}{e}\right)^{\ell}\ell^{\ell}(1+{\mathcal O}(\ell^{-1/2})).
\]
(Actually, the result in \cite{AgFuGiRo} was for normal galled trees; however, the same method of proof shows that the number of all galled trees has the same asymptotics.) Therefore, we obtain
\[
\vert\C\cap\N_n\vert\sim\frac{2^{2k-1}\sqrt{2e}}{(2k)!}n^{2k-1}\left(\frac{2}{e}\right)^nn^{n}
\]
and thus
\begin{equation}
\label{kg2}
    \lim_{n\rightarrow\infty}\frac{\vert\C\cap\N_n\vert}{\vert\N_n\vert}=\frac{2^kk!}{(2k)!}=\frac{1}{(2k-1)!!}>0\ \ \text{for}\ k\geq 2,
\end{equation}
which proves the claimed result for each $k \geq 2$.

It remains to consider the case $k=1$. Notice that, in this case, $\N=\C$ and thus, $\N$ is a closed class of networks, in contrast to the cases where $k>1$. 
Here, instead of $\C$, we consider the subclass $\C'$ of $\N$ consisting of all simplicial tree-child networks with at most $1$ reticulation. Then $\C'$ is a closed subclass of $\N$. Moreover it is clear that $\C' \neq \N$. Next, recall that it was proved in \cite{CaZh} that the number of simplicial tree-child networks $({\rm STC}_{\ell,k})$ with exactly $k$ reticulations and $\ell$ leaves that are labelled by the set $X=\{1,\ldots,\ell\}$ admits the closed-form expression:
\begin{equation}\label{STC}
{\rm STC}_{\ell,k}=\binom{\ell}{k}\frac{(2\ell-2)!}{2^{\ell-1}(\ell-k-1)!};
\end{equation}
Thus, by Stirling's formula, as $\ell\rightarrow\infty$,
\[
{\rm STC}_{\ell,k}=\frac{\sqrt{2}}{2k!}\ell^{2k-1}\left(\frac{2}{e}\right)^{\ell}\ell^{\ell}\left(1+{\mathcal O}(\ell^{-1})\right).
\]
and consequently for the number of simplicial tree-child networks $({\rm STC}_{\ell,\leq k})$ with at most $k$ reticulations,
\[
{\rm STC}_{\ell,\leq k}=\sum_{i=0}^{\ell}{\rm STC}_{\ell,i}=\frac{\sqrt{2}}{2k!}\ell^{2k-1}\left(\frac{2}{e}\right)^{\ell}\ell^{\ell}\left(1+{\mathcal O}(\ell^{-1})\right).
\]
Then, as above, where we set $k=1$:
\[
\vert\C'\cap\N_n\vert=\sum_{j=1}^{n}\binom{n}{j}{\rm STC}_{j,\leq 1}\sim\frac{\sqrt{2e}}{2}n\left(\frac{2}{e}\right)^{n}n^{n}
\]
which in turn yields
\[
\lim_{n\rightarrow\infty}\frac{\vert\C'\cap\N_n\vert}{\vert\N_n\vert}=\frac{1}{2}.
\]
This shows that $\N$ is not tight for $k=1$, too.
\end{proof}

\begin{rem}
Note that if $\C'$ denotes the class of simplicial-tree child networks with at most $k$ reticulations, then for all $k\geq 1$, we have
\[
\lim_{n\rightarrow\infty}\frac{\vert\C'\cap\N_n\vert}{\vert\N_n\vert}=2^{-k}.
\]
However, this does not show that $\N$ is not tight for $k\geq 2$ as $\C'$ is only a closed class for $k=1$. 
\end{rem}

Next, let $\N$ be the class of normal networks with at most $k$ (fixed) reticulations where again $k\geq 1$. By Lemma~\ref{largest-N}, the largest closed subclass of $\N$ is the set of phylogenetic trees, which we now denote by $\C$. Then, in contrast to tree-child networks, this class is tight.

\begin{prop}
The class $\N$ of normal networks with at most $k$ reticulations is tight for each $k\geq 1$.
\end{prop}
\begin{proof}
Normal networks with exactly $k$ reticulations and $\ell$ leaves that are labelled by the set $X=\{1,\ldots,\ell\}$ have also been counted in \cite{fuc}; see also \cite{FuStZh}. More precisely, denote the number of such networks by ${\rm NN}_{\ell,k}$. Then, it was shown in \cite{fuc,FuStZh} that ${\rm NN}_{\ell,k}$ admits the same first-order asymptotics as ${\rm TC}_{\ell,k}$:
\[
{\rm NN}_{\ell,k}=\frac{2^{k-1}\sqrt{2}}{k!}\ell^{2k-1}\left(\frac{2}{e}\right)^{\ell}\ell^{\ell}(1+{\mathcal O}(\ell^{-1/2}),\qquad (\ell\rightarrow\infty).
\]
Consequently, with the same proof as above, we have
\[
\vert\N_n\vert\sim\frac{2^{k-1}\sqrt{2e}}{k!}n^{2k-1}\left(\frac{2}{e}\right)^n n^{n}.
\]
Applying Eqn.~\ref{asym-Tn} to provide the upper bound on $\vert\C\cap\N_n\vert$, where $\C$ denotes the class of phylogenetic trees (see Lemma~\ref{largest-N}), and dividing by $|\N_n|$ shows that Eqn.~\ref{zero} holds for all $k\geq 1$. 
\end{proof}

\section{Concluding comments}
In this paper, we have introduced the notion of a class of networks being `tight' and shown that several standard classes of phylogenetic networks satisfy this property. These include the classes of trees, galled trees, tree-child networks, normal networks, and normal networks with at most $k$ reticulations for $k \geq 1$. On the other hand, we have also established that various other network classes fail to have this tightness property, in particular, the classes of galled networks, semi-simplicial networks, and  the class of tree-child networks with at most $k$ reticulations for $k \geq 1$. It may therefore be of interest in future work to consider which other classes of phylogenetic networks satisfying property ($P_2$) are tight.

\section{Acknowledgements}
We thank the reviewers for many useful comments. The second author thanks Andrew Francis for preliminary discussions regarding the concept of closed classes of networks and the tightness properties of trees. He also thanks the NZ Marsden Fund for research support (23-UOC-003). The first author acknowledges partial support by National
Science and Technology Council, Taiwan under research grant NSTC-113-2115-M-004-004-MY3.

\section{Appendix: Proof of Theorem ~\ref{ssn}}

In this appendix, we prove Theorem~\ref{ssn}. Note that semi-simplicial networks are galled networks which have been investigated in \cite{fuc22}. The proof of Theorem~\ref{ssn} relies heavily on tools from this paper and is similar to the proof of Eqn.~\ref{asymp-GNl}.

One crucial observation in \cite{fuc22} was the following: a galled network is almost surely a simplicial (galled) network $N$ with simplicial networks with two leaves attached to some of the leaves below reticulations of $N$. Thus, in order to count semi-simplicial networks, we only have to count simplicial networks $N$ with cherries attached to some of the leaves below reticulations of $N$. For this number ($\widetilde{{\rm SSN}}_\ell$), we have the following closed-form expression, where $N_{\ell+1}^{(k)}$ denotes the number of simplicial networks with $\ell$ leaves and $k$ reticulations, where the leaves below the reticulations have labels from the set $\{1,\ldots, k\}$ (this notation comes from \cite{fuc22}).

\begin{lem}
We have    
\begin{equation}\label{tildeSSN}
\widetilde{SSN}_{\ell}=\sum_{j=0}^{\lfloor \ell/2\rfloor}\binom{\ell}{2j}\frac{(2j)!}{2^jj!}\sum_{i=0}^{\ell-2j}\binom{\ell-2j}{i}N_{\ell-j+1}^{(i+j)}.
\end{equation}
\end{lem}
\begin{proof}
For $0\leq j\leq \ell/2$ and $0\leq i\leq\ell-j$, the number of simplicial networks with $i+j$ reticulations whose leaves are labelled by the set $\{1,\ldots,i+j\}$ and with $\ell-j$ leaves in total is given by $N_{\ell-j+1}^{(i+j)}$. For each such network, replace the leaves with labels $\{1,\ldots,j\}$ by cherries and then relabel all leaves so that the resulting labels are all different. For the relabelling, there are
\[
\binom{\ell}{2j}\binom{\ell-2j}{i}\frac{(2j)!}{2^jj!},
\]
ways, as we first have to choose $2j$ labels for the cherries and then have to separate the remaining $\ell-2j$ leaves into those which are below reticulations ($i$) and those which are not ($\ell-2j-i$). The last factor is because swapping the leaves of a cherry results in the same network and thus the $2j$ labels for the cherries have to be divided into $j$ disjoint sets of size $2$.
\end{proof}

\begin{proof}[Proof of Theorem~\ref{ssn}] As mentioned at the beginning of this appendix, we know from \cite{fuc22} that
\begin{equation}\label{SSNasymp}
{\rm SSN}_{\ell}\sim\widetilde{{\rm SSN}}_{\ell},\qquad (\ell\rightarrow\infty).
\end{equation}
Thus, we only need to concentrate on the latter sequence for which Eqn.~\ref{tildeSSN} holds. We break the first summation into two parts according to whether $j<\sqrt[4]{\ell}$ or $\sqrt[4]{\ell}\leq j\leq\lfloor\ell/2\rfloor$. The second part is shown to be an exponentially small function multiplied with $\ell^{2\ell}$ as in the proof of Proposition~4 in \cite{fuc22}. Thus, the main contribution comes from the  first part, for which we use the estimate of Lemma~10-(ii) from \cite{fuc22}, which reads:
\[
\sum_{i=0}^{\ell-2j}\binom{\ell-2j}{\ell}N_{\ell-j+1}^{(i+j)}=\frac{\sqrt{2\sqrt{e}}}{2^{3j+2}}\ell^{-1}\left(\frac{8}{e^2}\right)^{\ell}\ell^{2\ell-2j}\left(1+{\mathcal O}\left(\frac{j^2}{\ell}+\frac{1}{\sqrt{\ell}}\right)\right)
\]
uniformly for $j=o(\sqrt{\ell})$. Plugging this into the first part  gives:
\[
\widetilde{\rm SSN}_{\ell}\sim\left(\sum_{j<\sqrt[4]{\ell}}\frac{1}{j!16^j}\right)\frac{\sqrt{2\sqrt{e}}}{4}\ell^{-1}\left(\frac{8}{e^2}\right)^{\ell}\ell^{2\ell}\sim \frac{\sqrt{2\sqrt{e\sqrt[4]{e}}}}{4}\ell^{-1}\left(\frac{8}{e^2}\right)^{\ell}\ell^{2\ell}
\]
which together with Eqn.~\ref{SSNasymp} proves the claimed result.
\end{proof}
\end{document}